\documentclass{article} 
\usepackage{amsmath, amsthm}
\usepackage{amsfonts,amssymb}

\newcommand{\M}{\|}
\newcommand{\R}{\mathbb{R}}

\newcommand{\N}{\mathbb{N}}
\newcommand{\C}{\mathbb{C}}

\newcommand{\Cal}{\mathcal}

\newtheorem{Theorem}{Theorem}[section]

\newtheorem{Corollary}[Theorem]{Corollary}

\newtheorem{Lemma}[Theorem]{Lemma}
\newtheorem{Definition}[Theorem]{Definition}

\newtheorem{Remark}[Theorem]{Remark}

\renewcommand{\epsilon}{\varepsilon}

\begin{document}
\title{Reflectionless Sturm-Liouville Equations.}
\author{Robert Sims\\[10pt]
Department of Mathematics\\ University of California at Davis\\
Davis CA 95616, USA\\
Email: rjsims@math.ucdavis.edu}

\maketitle  
\begin{abstract}
We consider compactly supported perturbations of periodic 
Sturm-Liouville equations. In this context, one can use the Floquet
solutions of the periodic background to define 
scattering coefficients. We prove that if the reflection
coefficient is identically zero, then the operators corresponding to the
periodic and perturbed equations, respectively, are unitarily
equivalent. In some appendices, we also provide the proofs of 
several basic estimates, e.g. bounds and asymptotics for 
the relevant m-functions. 
\end{abstract}

\setcounter{section}{0}
\renewcommand{\theequation}{\arabic{section}.\arabic{equation}}
\newcounter{letters}


\section{Introduction}
Much can be understood about the properties of one dimensional
Sturm-Liouville equations by analyzing the corresponding 
reflection and transmission coefficients. The purpose of this article 
is to present some results concerning the implications of trivial
scattering, i.e., situations in which there is an absence of
reflection.

Throughout our work, we will consider equations of the form
\begin{equation} \label{1eq:bst}
-(pu')' + qu = \lambda u
\end{equation}
where $\lambda \in \C$ and the real-valued functions  $ \frac{1}{p}$
and $q$ are locally integrable with $p>0$ almost everywhere. As a
first example, consider equation (\ref{1eq:bst}) in the case that the
coefficients $1-p$ and $q$ are additionally assumed to have
compact support; take ${\rm supp}(1-p) \subset [0,1]$ and ${\rm supp}(q) \subset [0,1]$
for simplicity. For such equations, 
one defines the classical scattering coefficients by examining
the Jost solution of (\ref{1eq:bst}) which satisfies 
\begin{equation} \label{1eq:Jost}
u(x, k) = \left\{ \begin{array}{cc} e^{ikx} & \mbox{for } x \leq 0, 
\\ a(k) e^{ikx} + b(k) e^{-ikx} & \mbox{for } x \geq 1. \end{array} \right.
\end{equation}
for $0<k$ with $\lambda = k^2$. The coefficients $a(k)$ and
$b(k)$ describe the reflected and transmitted parts of an incoming plane 
wave, a solution of the free equation, and in the
physics literature one often defines  the transmission
coefficient by $t(k) := \frac{1}{a(k)}$ and the reflection
coefficient by $r(k) := \frac{b(k)}{a(k)}$. In our work, we find it more
useful to deal with the coefficients $a(k)$ and $b(k)$ directly, and
we label them the transmission and reflection coefficient,
respectively. We will be specifically interested in cases where the
equation (\ref{1eq:bst}) is {\em reflectionless}, i.e., in the event that 
$b(k) =0$ for all $k>0$. 

Let us briefly recall some of the known results. 
In the specific case mentioned above, in which both 
$1-p$ and $q$ are supported in $[0,1]$, one 
can consider the equation
\begin{equation} \label{1eq:perext}
- ( \tilde{p}u')' + \tilde{q}u = \lambda u
\end{equation}
where $\tilde{p}$ and $\tilde{q}$ are the 1-periodic extensions of $p$
and $q$, respectively, to $\R$. It is easy to see, e.g. 
Lemma 3.1 of \cite{SS}, that if equation (\ref{1eq:bst}) is reflectionless, 
then the periodic operator corresponding to (\ref{1eq:perext}) has gapless
spectrum equal to $[0, \infty)$. If one assumes in addition that $p
\equiv 1$, then a result due to Borg 
\cite{Borg1, Borg2} for continuous $q$, later extended to 
integrable $q$ by Hochstadt \cite{H}, may be used to prove that
reflectionless, i.e., gapless spectrum for the periodic operator, 
implies $q \equiv 0$. Such results correspond to the well known fact 
that there are no compactly supported solitons.

If $p$ is sufficiently smooth, e.g. $p$ and $p'$ are absolutely
continuous, then one may apply a unitary, Liouville-Green 
transformation, as is done in \cite{Everitt}, and see that equation (\ref{1eq:bst}) is 
equivalent to a Schr\"odinger equation of the form $-u'' + Q u =
\lambda u$, where $Q$ is a function involving both $p$ and $q$. Based
on these observations, it was proven in Theorem 3.3 of \cite{SS} that if
$1-p$ and $q$ have support $[0,D]$ with $p$ and $p'$
absolutely continuous, then (\ref{1eq:bst}) is reflectionless if and
only if 
\begin{equation} \label{1eq:q=f(p)}
q = \frac{1}{16} \frac{(p')^2}{p} - \frac{1}{4} p''.
\end{equation} 
Using this equation, one can readily see that if $q \equiv 0$, $1-p$ is
compactly supported, $p$ and $p'$ are absolutely continuous, and
equation (\ref{1eq:bst}) is reflectionless, then $p \equiv
1$. Moreover, one may establish that such a result remains true with no
additional smoothness assumptions required on $p$. Specifically,
Proposition 4.1 of \cite{SS} states that if $q \equiv 0$, $1-p$ is
compactly supported, and (\ref{1eq:bst}) is reflectionless then $p
\equiv 1$. This result was proven by examining 
the asymptotics of the $m$-function corresponding to (\ref{1eq:bst}).
In partcular, it did not use known results for the Schr\"odinger
equation due to the fact that the classical Liouville-Green 
operator is ill-defined when $p$ is not smooth.   

In this article, we will generalize some of these results to 
reflectionless equations whose scattering coefficients are defined 
with respect to a periodic background. More specifically, we consider 
1-periodic, real-valued functions $p_0$ and $q_0$ for which both $\frac{1}{p_0}$
and $q_0$ are locally integrable with $p_0 >0$ almost everywhere. It
is well known, see e.g. \cite{Eastham, CL, Weidmann}, that the operator
\begin{equation} \label{1eq:H0}
H_0 = - \frac{d}{dx} p_0 \frac{d}{dx} + q_0
\end{equation}
on $L^2( \R)$ has spectrum which consists of a union of bands.
For $\lambda$ in a stability interval, 
there exist linearly independent solutions of  
\begin{equation} \label{1eq:perst}
-(p_0u')' + q_0 u = \lambda u,
\end{equation} 
which we label by $\phi_{\pm}( \cdot, \lambda)$ and refer to as the {\em Floquet solutions} corresponding to
(\ref{1eq:perst}), see Section~\ref{sec:perscat} for a further discussion. 
Let $f \geq 0$ and $g$ be real-valued, integrable
functions with compact support contained in $[0, \infty)$. 
Define perturbations of the periodic coefficients
introduced above by setting $\frac{1}{p} :=  \frac{1}{p_0}+f$ and
 $q := q_0 +g$. The {\em periodic scattering coefficients} are defined in
 terms of the solution of
\begin{equation} \label{1eq:pertper}
-(pu')' + qu = \lambda u
\end{equation} 
which satisfies
\begin{equation} \label{1eq:Jost+}
u_+(x, \lambda) = \left\{ \begin{array}{cc} \phi_+(x, \lambda) & \mbox{for } x \leq 0, 
\\ a_p( \lambda) \phi_+(x, \lambda) + b_p( \lambda) \phi_-(x, \lambda) & \mbox{for } x \geq D, \end{array} \right.
\end{equation} 
again, for $\lambda$ in a stability interval. Here $D>0$ is chosen as
$\inf \{ D'>0 \, : \, \mbox{supp}(f) \, \cup \, \mbox{supp}(g) \subset [0, D'] \}$. Comparing with (\ref{1eq:Jost}), we
see that the Floquet solutions play the role of 
the plane waves when the periodic background is non-trivial. 
Analogously to equation (\ref{1eq:H0}), we will denote by $H$ the operator on 
$L^2( \R)$ corresponding to the coefficients $p$ and $q$. We say that
equation (\ref{1eq:pertper}) (the operator $H$) is {\em reflectionless} with respect to
(\ref{1eq:perst}) (the operator $H_0$) if there exists a non-empty
stability interval in the spectrum of $H_0$ on
which $b_p( \lambda)$ is identically zero. We note that due to the analyticity
of $b_p$, see Lemma~\ref{lem:analytic} below, it is sufficient to 
assume that there exists a stability interval on which the zeros of 
$b_p$ have an accumulation point.    

We prove the following theorem.
\begin{Theorem} \label{thm:uni} If $H$ is reflectionless with respect
to $H_0$, then $H$ is unitarily equivalent to $H_0$.
\end{Theorem}

Our proof follows from a result of Bennewitz \cite{Ben2} which establishes
the existence of a more general Liouville Transform. Moreover, since
this unitary map is explicit, we may also prove
\begin{Corollary} \label{cor:triv} Let $p_0$, $q_0$, $f$, and $g$ be given
  as above. Suppose that $f \equiv 0$ and $H$ is 
reflectionless with respect to $H_0$, then $g \equiv 0$.
\end{Corollary}
 
The paper is organized as follows. In Section 2, we introduce
scattering coefficients defined with respect to a periodic background.
Next, we prove Theorem~\ref{thm:uni} and Corollary~\ref{cor:triv} in
Section 3. In the appendices that follow, we provide a proof of the
technical estimates necessary to apply the inverse results found in
\cite{Ben2}. The first appendix, Appendix~\ref{App:apri}, establishes 
some bounds on the growth of solutions to Sturm-Liouville equations. 
A convergence result for the corresponding $m$-functions is given in
Appendix~\ref{App:mfun}. Appendix~\ref{App:mfunasy} contains 
the main results which describe the asymptotics of the $m$-function 
and thereby the Weyl solution.

{\em Acknowledgements:} The author wishes to thank the referee for
the many useful comments made on a previous version of this
paper. Moreover, the author is indebted to both G\"unter Stolz and Bruno
Nachtergaele for many stimulating discussions. 

\setcounter{equation}{0}


\section{Scattering at a Periodic Background} \label{sec:perscat}

In this section, we recall some of the basic facts concerning periodic
Sturm-Liouville equations. For a more detailed discussion of periodic
problems, we refer the reader to \cite{Eastham}. We also note that 
many of the results stated below are proven, for Schr\"odinger 
equations, in \cite{DSS}.

\subsection{The Floquet Solutions}
Let $p_0$ and $q_0$ be 1-periodic, real valued functions for which 
both $\frac{1}{p_0}$ and $q_0$ are locally integrable and $p_0>0$
almost everywhere. Consider the
self-adjoint operator on $L^2( \R)$ given by
\begin{equation} \label{2eq:H0}
H_0 \, := \, - \frac{d}{dx}p_0 \frac{d}{dx} \, + \, q_0.
\end{equation}
The domain of this operator is the set of all $f \in L^2( \R)$ for
which both $f$ and $p_0f'$ are absolutely continuous and $H_0f \in L^2(
\R)$. Since $p_0$ is not assumed to be smooth, in this generality we
allow $p_0$ to be a step function for example, we note that the smoothness
of $p_0f'$ is not necessarily inherited by $f'$. For any 
$z \in \C$, let $u_N( \cdot, z)$ and $u_D( \cdot, z)$ denote the solutions of
\begin{equation} \label{2eq:perbst}
-(p_0u')' + q_0u = zu,
\end{equation}
satisfying
\begin{equation} \label{2eq:norm}
\left( \begin{array}{cc} u_N(0,z) & u_D(0,z) \\
p_0u'_N(0,z) & p_0u'_D(0,z) \end{array} \right) \, = \, I.
\end{equation}
Take $g_0(z)$ to be the transfer matrix of (\ref{2eq:perbst}) from $x=0$ to
$x=1$, i.e., the matrix for which 
\begin{equation} \label{2eq:tran}
\left( \begin{array}{c} u(1,z) \\ p_0 u'(1,z) \end{array} \right) = g_0(z)
\left( \begin{array}{c} u(0,z) \\ p_0 u'(0,z) \end{array} \right), 
\end{equation}
for any solution $u$ of (\ref{2eq:perbst}). Set $\rho_{\pm}(z)$ to be the
eigenvalues of $g_0(z)$, i.e., the roots of $\rho^2 -D(Z) \rho
+1=0$, where $D(z)= \mbox{Tr}[g_0(z)]$. The spectrum of $H_0$ 
consists of bands which are given by the sets of real numbers 
$E$ for which $|D(E)| \leq 2$. A {\em stability interval} of 
$H_0$ is a maximal interval, $(c,d)$, such that $|D(E)| < 2$ for every
$E \in (c,d)$. It is on such an interval, and appropriate
analytic extensions thereof, that one may define the Floquet
solutions. We state these results as a lemma.

Let $p_0$ and $q_0$ be as above and fix a stability interval 
$(c,d)$ in the spectrum of the operator $H_0$. 
Consider the open vertical strip in the complex plane 
containing $(c,d)$, i.e.,
\begin{equation} \label{2eq:strip}
S_{(c,d)} \, := \, \{ z \in \C : \, z = E + i \eta, \ \ \mbox{where} \ \ c
< E < d \ \ \mbox {and} \ \ \eta \in \R \}.
\end{equation}

\begin{Lemma}\label{lem:Fprelim} Let $p_0$, $q_0$, and $S_{(c,d)}$ be taken as
  above. \\
\emph{i)} As functions of $z$, the eigenvalues $\rho_{\pm}$ of $g_0$
may be chosen analytic in $S_{(c,d)}$ with, at most, algebraic 
singularities at the points $z=c$ and $z=d$. \\
\emph{ii)} For the choices of $\rho_{\pm}$ taken in i) above, one may
define eigenvectors $v_{\pm}$ of $g_0$, corresponding to
$\rho_{\pm}$, that are analytic in $S_{(c,d)}$ with, at most,
algebraic singularities at $c$ and $d$.
\end{Lemma}

Using Lemma~\ref{lem:Fprelim}, one defines the Floquet solutions 
to be the solutions of (\ref{2eq:perbst}) which satisfy the initial
conditions 
\begin{equation} \label{2eq:floqnorm}
\left( \begin{array}{c} \phi_{\pm}(0,z) \\ p_0 \phi'_{\pm}(0,z)
  \end{array} \right) = v_{\pm}(z),
\end{equation}
for any $z \in S_{(c,d)}$. We state the properties of these 
solutions as a separate lemma.

\begin{Lemma} \label{lem:Floq} Let $p_0$, $q_0$, and $S_{(c,d)}$ be as
  above, and define the Floquet solutions $\phi_{\pm}( \cdot, z)$ as
  in (\ref{2eq:floqnorm}) above. We have that \\ 
\emph{i)} For any fixed $x$, both the solutions $\phi_{\pm}(x, \cdot)$
and the corresponding derivatives $p_0 \phi'_{\pm}(x, \cdot)$ are 
analytic in $S_{(c,d)}$ with, at most, algebraic 
singularities at the points $z=c$ and $z=d$. \\
\emph{ii)}  For every fixed $z \in S_{(c,d)}$, the set 
$\{ \phi_+( \cdot, z), \phi_-( \cdot, z) \}$ constitutes a 
basis for the solution space corresponding to (\ref{2eq:perbst}). \\
\emph{iii)} Upon labeling the eigenvalues $\rho_{\pm}$ appropriately, 
for any $E \in (c,d)$, one has that $\phi_{\pm}( \cdot, E+i \eta) \in
L^2$ near $\pm \infty$ for $\eta >0$ and similarly $\phi_{\pm}( \cdot, E+i \eta) \in
L^2$ near $\mp \infty$ for $\eta <0$.
\end{Lemma}

The proofs of both Lemma~\ref{lem:Fprelim} and Lemma~\ref{lem:Floq} 
are provided in Section 2.1 of \cite{DSS} in the
context of Schr\"odinger equations. Substituting the definitions given
above, one may easily translate these proofs to the Sturm-Liouville
equations we consider here.

\subsection{Periodic Scattering Coefficients}
Fix $p_0$, $q_0$, and $S_{(c,d)}$ as defined in the previous
subsection. As in the introduction, let $f \geq 0$ and $g$ be 
real-valued, integrable functions whose supports are contained in $[0,D]$. 
Define perturbations of $p_0$ and $q_0$ by the equations 
$\frac{1}{p}  :=  \frac{1}{p_0}+f$ and  $q  :=  q_0 +g$,
respectively. For any $z \in S_{(c,d)}$, let $u_+$ be the solution of
\begin{equation} \label{2eq:pertbst}
-(pu')' + qu =zu
\end{equation}
satisfying
\begin{equation} \label{2eq:perJost}
u_+(x,z) \,  := \, \left\{ \begin{array}{cc} \phi_+(x,z) & \mbox{for } x
\leq 0 \\ a_p(z) \phi_+(x,z) + b_p(z) \phi_-(x,z) & \mbox{for }
x \geq D. \end{array} \right.
\end{equation}
As indicated by Lemma~\ref{lem:Floq} ii), 
the Floquet solutions are linearly independent for $z \in S_{(c,d)}$, and
therefore, $a_p(z)$ and $b_p(z)$ are uniquely defined. 
In this setting, $b_p$ and $u_+$ take on the role of 
a modified reflection coefficient and Jost solution, respectively, 
relative to the periodic background. 

\begin{Lemma} \label{lem:analytic}Let $(c,d)$ be a stability interval of $H_0$. 
The scattering coefficients $a_p( \cdot)$ and 
$b_p( \cdot)$ are analytic for $z \in S_{(c,d)}$ with, at most,
algebraic singularities at the points $z=c$ and $z=d$.
\end{Lemma}

\begin{proof} 
Let $g_1(z)$ denote the transfer matrix corresponding to
(\ref{2eq:pertbst}) from $x=0$ to $x=D$, in analogy with (\ref{2eq:tran}). Clearly then,
\begin{equation} \label{2eq:u+at1}
\left( \begin{array}{c} u_+(D,z) \\ pu_+'(D,z) \end{array} \right) =
g_1(z) \left( \begin{array}{c} \phi_+(0,z) \\ p_0 \phi_+'(0,z) \end{array} \right).
\end{equation}
It is well known that these solutions and their derivatives are, at fixed $x$, 
analytic in $z$, and therefore, the entries of $g_1(z)$ are entire as functions of $z$.
Using Lemma~\ref{lem:Floq} i), we conclude that the left hand
side of (\ref{2eq:u+at1}) is analytic in $S_{(c,d)}$ with, at most,
algebraic singularities at the points $z=c$ and $z=d$. 
From (\ref{2eq:perJost}), one sees that 
\begin{equation}
\left( \begin{array}{c} a_p(z) \\ b_p(z) \end{array} \right) \, = \, 
\left( \begin{array}{cc} \phi_+(D,z) & \phi_-(D,z)  \\  p \phi_+'(D,z)
  & p \phi_-'(D,z) \end{array} \right)^{-1} \,
\left( \begin{array}{c} u_+(D,z) \\ pu_+'(D,z) \end{array} \right),
\end{equation}
by which this lemma follows from another application of Lemma~\ref{lem:Floq}.
\end{proof}
In what follows, we will denote by $H$ the 
self-adjoint operator in $L^2( \R)$ 
corresponding to equation (\ref{2eq:pertbst}) 
in analogy to (\ref{2eq:H0}). We will say that equation (\ref{2eq:pertbst}) (the
operator $H$) is {\em reflectionless} with respect to equation (\ref{2eq:perbst}) (the
operator $H_0$) if there is a stability interval $(c,d)$ in the spectrum of
$H_0$ for which $b_p( \lambda)=0$ for all $\lambda \in (c,d)$; thus,
$b_p(z) = 0$ for all $z \in S_{(c,d)}$ by Lemma~\ref{lem:analytic} above.

\setcounter{equation}{0}


\section{Proofs of the Main Results} \label{sec:proofs}
In this section, we will prove Theorem~\ref{thm:uni} and 
Corollary~\ref{cor:triv}. Our proofs are based on a 
recent inverse result of Bennewitz \cite{Ben2} in which he constructs
a more general Liouville transform; specifically, one that is 
applicable even in the case of non-smooth $p$. We will begin 
by describing his transformations and then verify that
we may apply his inverse results. 



\subsection{Liouville Transforms}
Bennewitz's results apply to half-line operators. 
To state his main theorem, let $p$ and $q$ be real-valued functions on 
$[0, \infty)$ with both $\frac{1}{p}$ and $q$ in $L^1_{\rm loc}( 0,
\infty)$. Consider the Sturm-Liouville equation
\begin{equation} \label{3eq:bst}
-(pu')' + qu = \lambda u,
\end{equation}
subject to the boundary condition 
\begin{equation} \label{3eq:bstbc}
u(0) \cos( \alpha) +pu'(0) \sin( \alpha) = 0,
\end{equation}
for some $\alpha \in [0, \pi)$. Bennewitz's results are applicable
under rather general assumptions. For the results we wish to present,
we will assume more than is necessary as we indicate briefly below. 

\noindent i) We assume that we are working on a half-line; the results
also apply in the case that $p$ and $q$ are as above, yet 
defined on $[0,b)$ with $b< \infty$.
\newline ii) We assume that $p >0$ almost everywhere. Technically, one
need only assume $\frac{1}{p} \in L^1_{\rm loc}$.
\newline iii) We will assume that equation (\ref{3eq:bst}) is
limit-point at $+ \infty$, see Chapter 9 of \cite{CL} for details.

\begin{Definition} \label{def:lt}
A unitary Liouville transform is a map $F: L^2( 0, \infty) \to L^2( 0,
\infty)$ satisfying
\begin{equation} \label{3eq:lgt}
(Fv)(x) = s(x) v(t(x)),
\end{equation}
where $s \in L^2_{loc}(0, \infty)$ is such that $s \neq 0$ almost
everywhere, $t(x)= \int_0^x|s(y)|^2dy$, and $ \lim_{b \to \infty}t(b)
=  \infty$.
\end{Definition}
One may easily check that such a map is unitary and that $F^{-1}$,
the inverse of $F$, is also a unitary Liouville transform. 

Now, for $i=1,2$, let $p_i$ and $q_i$ be functions which satisfy the
conditions stated above and let $\alpha_i \in [0, \pi)$. 
Denote by $H_i$ the self-adjoint operator
on $L^2(0, \infty)$ generated by equation (\ref{3eq:bst}) and 
(\ref{3eq:bstbc}) with coefficients $p_i$ and $q_i$ and boundary
conditions $\alpha_i \in [0, \pi)$. We use the following 
result from \cite{Ben2}:

\begin{Theorem} \label{thm:Ben}
Suppose the operators $H_1$ and $H_2$ have the same 
spectral measure. Then there is a unitary Liouville 
transform $F$ mapping $H_2$ to $H_1$; specifically, $FH_2=H_1F$.
\end{Theorem}
Bennewitz explicitly constructs this unitary Liouville transformation 
in terms of the mappings $t_i:[0, \infty) \to [0, \infty)$ defined by
\begin{equation} \label{3eq:ti}
t_i(x) \, := \, \int_0^x \left( \frac{1}{p_i(y)} \right)^{1/2}dy.
\end{equation}
He then defines $t:[0, \infty) \to [0, \infty)$ as $t(x) =
t_2^{-1}(t_1(x))$ and subsequently, for almost every $x \in [0, \infty)$,
\begin{equation} \label{3eq:s}
s(x) \, := \, \sqrt{t'(x)} \, = \, \left( \frac{p_2(t(x))}{p_1(x)}
\right)^{1/4} \, > \, 0.
\end{equation}

We note that although the spectral measures of $H_1$
and $H_2$ depend on the boundary conditions $\alpha_1$ and $\alpha_2$,
respectively, the functions $s$ and $t$, and therefore the Liouville 
transform $F$, do not.



\subsection{Proofs}

We may now provide the proofs of Theorem~\ref{thm:uni} and
Corollary~\ref{cor:triv}. The proof uses well known 
results concerning the $m$-function corresponding to the
Sturm-Liouville equations we are considering. We refer the interested
reader to Chapter 9 of \cite{CL} for a complete discussion.  

\noindent {\em Proof of Theorem~\ref{thm:uni}} 
If $H$ is reflectionless with respect to $H_0$, then there exists a
stability interval $(c,d)$ in the spectrum of $H_0$ for which $b_p(
\lambda) = 0$ for all $\lambda \in (c,d)$. As $b_p$ is analytic on
$S_{(c,d)}$, we conclude that $b_p(z) = 0$ for all $z \in
S_{(c,d)}$. Using (\ref{2eq:perJost}), we see that for $\lambda \in (c,d)$ and
$\eta >0$ the modified Jost solution satisfies 
\begin{equation}
u_+(x, \lambda + i \eta) = a_p( \lambda + i \eta) \phi_+(x, \lambda +
i \eta),
\end{equation}
for all $x \geq D$. By Lemma~\ref{lem:Floq} iii), the Floquet solution $\phi_+(
\cdot, \lambda + i \eta)$ is square integrable at $+ \infty$. Thus,
the modified Jost solution $u_+$ coincides, up to a complex multiple,
with the Weyl solution. Appealing again to (\ref{2eq:perJost}), it is
clear that for $\lambda + i \eta \in S_{(c,d)}$ with $ \eta>0$, 
the $m$-function for the perturbed equation (\ref{1eq:pertper}) satisfies 
\begin{equation} \label{3eq:m=m}
m(0, \lambda + i \eta) = \frac{pu_+'(0, \lambda + i \eta)}{ u_+(0,
  \lambda + i \eta)} = \frac{p_0 \phi_+'(0, \lambda + i \eta)}{
  \phi_+(0, \lambda + i \eta)} = m_0(0, \lambda + i \eta),
\end{equation}
where $m_0$ is the corresponding $m$-function for the periodic
equation (\ref{1eq:perst}). As the $m$-functions are 
analytic on the upper half plane, the equality in (\ref{3eq:m=m}) 
holds throughout the upper half plane. From equality of the 
$m$-functions, we can conclude the equality of the spectral measures
of the half-line operators, with Dirichlet boundary condition at
$x=0$, corresponding to equations (\ref{1eq:perst}) and (\ref{1eq:pertper}), respectively.
Applying Bennewitz's result Theorem~\ref{thm:Ben}, we find an explicit
unitary equivalence of the Dirichlet operators on $[0, \infty)$. 
Since the coefficients of (\ref{1eq:perst}) and (\ref{1eq:pertper})
are identical on $(- \infty, 0]$, this unitary transformation, which
does not depend on the boundary condition at $x=0$, can be extended by
the identity to a unitray transformation corresponding to the whole
line. We have proven the theorem.  \hfill \qed

\vspace{.3cm}

\noindent {\em Proof of Corollary~\ref{cor:triv}:}
Suppose $f \equiv 0$, i.e., $p \equiv p_0$ and $H$ is reflectionless
with respect to $H_0$. In this case, it is clear
from (\ref{3eq:ti}) that the mappings $t_1 \equiv t_2$, for this application
we take $t_1$ to be defined in terms of $p_0$ and $t_2$ in terms of $p$, 
and therefore $t(x) = x$. Using (\ref{3eq:s}), we see that $s \equiv
1$, and therefore, the unitary Liouville transform $F$ is the
identity. From this, the equation $F H = H_0 F$ implies that $q_0(x) =
q(x)$, i.e. $g \equiv 0$.  \hfill \qed

\vspace{1cm} 

\setcounter{equation}{0}


\renewcommand{\theequation}{\thesection.\arabic{equation}}

\begin{appendix}

\begin{center}
\Large {\bf Appendix}
\end{center}

\vspace{.5cm}

\normalsize

The goal of the appendices that follow is to prove the key technical
estimate, Lemma 2.4 of \cite{Ben2}, which enables Bennewitz to prove
Theorem~\ref{thm:Ben}. Such estimates were originally proven by Bennewitz 
in \cite{Ben1}, see specifically Theorem 6.1 and Corollary 6.2, and they
are applicable under rather general assumptions; for example, the
coefficients of the basic Sturm-Liouville equation may be taken to be
measures. Our approach is more pedestrian, in particular, we assume
the coefficients are in $L^1_{\rm loc}( \R)$, but we hope our
streamlined presentation is more easily accessible. 

The heart of the matter is contained in Appendix~\ref{App:mfunasy} 
where we prove Theorem~\ref{Cthm:solasy}, the analogue of Lemma 2.4 of
\cite{Ben2}. The proof of Theorem~\ref{Cthm:solasy} uses a convergence result for
$m$-functions which are defined with respect to a sequence of
Sturm-Liouville equations whose coefficients converge in $L^1_{\rm
  loc}(\R)$. In Appendix~\ref{App:mfun}, we prove Lemma~\ref{Blem:mbds} and
Theorem~\ref{Bthm:mc} which demonstrate the desired convergence of the
$m$-functions. These results are new. Moreover, as is discussed in the
appendix of \cite{KKS} for equations with coefficients in $L^2_{\rm loc}( \R)$, they 
constitute a generalization of the applicability of Kotani Theory, see
\cite{K1, K2, K3}, to the Sturm-Liouville equations considered here. Lastly, all
the results presented in these appendices rely heavily on certain
basic solution estimates. The proofs of these results, which are
simple generalizations of estimates well known in the context of
Schrodinger equations, i.e., when $p \equiv 1$, are presented in
Appendix~\ref{App:apri}.


\section{A Priori Solution Estimates} \label{App:apri}

In this first appendix, we provide several standard solution estimates which
we will use frequently in the appendices that follow. Although results of this type are
well-known, see e.g. \cite{CL, PF, Weidmann}, we include the proofs here for the
convenience of the reader.

The basic Sturm-Liouville equation we consider is
\begin{equation} \label{Aeq:sleq}
-(pu')' +qu = 0,\end{equation}
where it is assumed throughout that $\frac{1}{p} \in L^1_{loc}(\R)$,
$p>0$ almost everywhere, and $q \in
L^1_{loc}(\R)$ may be complex valued. In most of our applications, we 
have $q = q_0 - \lambda$ for some real valued, locally integrable 
function $q_0$ and $\lambda \in \C$ a constant. For any 
$f \in L^1_{loc}(\R)$ and $I \subset \R$, a bounded interval, we will denote by 
\begin{equation} \label{Aeq:normf}
\M f \M_I \, := \, \int_I|f(t)|dt.
\end{equation}

\begin{Lemma} \label{Alem:gb}
Let $u$ be a solution of (\ref{Aeq:sleq}). For any $x,y \in \R$, 
we have that
\begin{equation} \label{Aeq:solbd}
| u(x)| ^2 + |pu'(x)|^2 \leq  \left( | u(y)| ^2 + |pu'(y)|^2 \right)  \exp
\left( \left\M \frac{1}{p}+q  \right\M_I \right),
\end{equation}
where the interval $I:= [ \min(x,y), \max(x,y)]$.
\end{Lemma}

\begin{proof} Setting $R(t) := |u(t)|^2+|pu'(t)|^2$, one easily calculates that 
\begin{equation} \label{Aeq:derR}
|R'(t)| = \left| 2{\rm Re}\left[ \left( \frac{1}{p(t)}+q(t) \right) u(t)\overline{pu'(t)} \right] \right| \le \left|
\frac{1}{p(t)}+q(t) \right| R(t).
\end{equation}  
Thus, $|(\ln R(t))'| \le |\frac{1}{p(t)}+q(t)|$, and the lemma
is proven.
\end{proof}

\begin{Lemma} \label{Alem:db}
For $i=1,2$, let $p_i$ and $q_i$ be functions with 
$\frac{1}{p_i} \in L^1_{loc}(\R)$, $p_i >0$ almost everywhere, and 
$q_i \in L^1_{loc}(\R)$. Suppose $u_i$ are solutions of
$-(p_iu_i')'+q_iu_i = 0$ which satisfy $u_1(y) = u_2(y)$ and
$p_1u_1'(y) = p_2u_2'(y)$ for some $y \in \R$. Then, for any $x \in
\R$ there exists a constant $C>0$ for which 
\begin{eqnarray} \label{Aeq:solest}
\left( | u_1(x) - u_2(x) | ^2 + |p_1u_1'(x) - p_2u_2'(x)|^2 \right)^{1/2}
\quad \quad \quad   \\ \leq \, C \, \left( | u_1(y)| ^2 + |p_1u_1'(y)|^2 \right)^{1/2} \exp
\left( \sum_{i=1}^2 \, \left\M \frac{1}{p_i}+ |q_i| \right\M_I \right), \nonumber
\end{eqnarray}
and one may take
\begin{equation} \label{Aeq:fatI}
C^2  = \left\M \frac{1}{p_1} - \frac{1}{p_2} \right\M^2_I + \M
    q_1-q_2 \M_I^2.
\end{equation}
Here $I:= [ \min(x,y), \max(x,y)]$. 
\end{Lemma}

\begin{proof} Without loss of generality, we consider the case of $y
  \le x$. For $i = 1,2$, define the vector 
\begin{equation}
\vec{u}_i(t) := \left( \begin{array}{c} u_i(t) \\
 p_iu_i'(t) \end{array} \right), 
\end{equation} 
for any $t \in \R$. Using this notation, the solutions 
$u_1$ and $u_2$ clearly satisfy
\begin{eqnarray} \label{Aeq:soldif}
\vec{u}_1(s) - \vec{u}_2(s) & = & \int_y^s \left(
  \begin{array}{c} \left( \frac{1}{p_1(t)} - \frac{1}{p_2(t)}
    \right)p_1u_1'(t) \nonumber \\
(q_1(t)-q_2(t)) u_1(t) \end{array} \right) \,dt +
\\ \mbox{ } & + & \int_y^s \left( \begin{array}{cc} 0 & \frac{1}{p_2(t)} \\ q_2(t) & 0 \end{array} \right) \left(
\vec{u}_1(t)-\vec{u}_2(t) \right) \,dt, 
\end{eqnarray}
for any  $y \le s \le x$. With the usual vector norm $\M \cdot \M$, one may
estimate that
\begin{multline} \label{Aeq:solbound}
\M \vec{u}_1(s)- \vec{u}_2(s) \M \, \leq \, \max \left\{
    \sup_{t \in I}|u_1(t)|, \sup_{t \in I}|p_1u_1'(t)|
  \right\} \, C \, + \\ + \int_y^s \left(
\frac{1}{p_2(t)} + |q_2(t)| \right) \M \vec{u}_1(t)-
\vec{u}_2(t) \M \,dt.
\end{multline}
By Lemma \ref{Alem:gb}, we may conclude that for $\psi \in
\{ u_1, p_1u_1' \}$, 
\begin{equation} \label{Aeq:psibo} 
\sup_{t \in I} |\psi(t)|  \leq \exp \left( \frac{1}{2} \left\M
    \frac{1}{p_1}+q_1 \right\M_I \right) \M
\vec{u}_1(y) \M.
\end{equation}
An application of Gronwall's lemma, see e.g. \cite{Walter}, to 
inequality (\ref{Aeq:solbound}) yields (\ref{Aeq:solest}) as desired.
\end{proof}

In the next lemmas we will provide local estimates from below on the
average growth of solutions to equation (\ref{Aeq:sleq}). For any
function $f \in L^1_{{\rm loc}}(\R)$ and any compact interval $I
\subset \R$, we will denote by 
\begin{equation} \label{Aeq:uniloc+}
\| f \|_{I, {\rm loc}+} \, := \, \sup_{x \in I} \, \int_x^{x + 1} |f(t)| dt \, < \, \infty,
\end{equation}
and 
\begin{equation} \label{Aeq:uniloc-}
\| f \|_{I, {\rm loc}-} \, := \, \inf_{x \in I} \, \int_x^{x+ 1} |f(t)| dt \, \geq \, 0.
\end{equation}

\begin{Remark} Since we assume throughout that $p>0$ and 
$\frac{1}{p} \in L^1_{{\rm loc}}( \R)$, for any compact (non-empty)
interval $I \subset \R$, the function $P: I \to (0, \infty)$ defined
by
\begin{equation}
P(x) \, := \, \int_x^{x+1} \frac{1}{p(t)} dt, 
\end{equation} 
is continuous; hence $\| \frac{1}{p} \|_{I, {\rm loc}-} >0$.
\end{Remark}

\begin{Lemma} \label{Alem:L2b}
Let $p$ and $q$ be functions with $\frac{1}{p} \in L^{1}_{ {\rm loc}}(
\R)$, $p>0$ almost everywhere, and $q \in L^{1}_{ {\rm loc}}( \R)$. 
For any (non-empty) interval $I =[a,b] \subset
\R$, there exists $C>0$ such that for all
real valued solutions of $-(pu')' + qu = 0$ and any $x \in I$,
\begin{equation} \label{Aeq:intbd}
\int_x^{x+2} |u(t)|^2 dt \geq C \left( |u(x)|^2 + | pu'(x)|^2 \right).
\end{equation}
\end{Lemma}

\begin{proof} Fix $x \in I$ and set $\tilde{I} = [a,b+2]$. By Lemma \ref{Alem:gb}, there are constants
$0<C_1, C_2<\infty$, depending only on $\M \frac{1}{p} \M_{ \tilde{I}, {\rm loc}+}$ and $\M q
\M_{\tilde{I}, {\rm loc}+}$, such that any solution of  $-(pu')' + qu = 0$ satisfies 
\begin{equation} \label{Aeq:amp}
C_1 \, \left( |u(x)|^2 + |pu'(x)|^2 \right) \, \leq \, |u(t)|^2 +
|pu'(t)|^2 \, \leq \, C_2  \left( |u(x)|^2 + |pu'(x)|^2 \right),
\end{equation}
for all $t \in [x,x+2]$. With $C_3 := (C_1/2)^{1/2}$ and $C_4 := (2C_2)^{1/2}$, 
we also have that 
\begin{equation} \label{Aeq:sum}
C_3 \, \left( |u(x)| + |pu'(x)| \right) \, \leq \, |u(t)| +
|pu'(t)| \, \leq \, C_4  \left( |u(x)| + |pu'(x)| \right).
\end{equation}

We now claim that there exists an $x_0 \in [x+ 1/2, x+ 3/2 ]$ for
which 
\begin{equation} \label{Aeq:uatx0}
|u(x_0) | \geq  \frac{C_3}{4}  \min \left(
\left\M \frac{1}{p} \right\M_{\tilde{I},{\rm loc}-}, 1 \right) \left( |u(x)| + |pu'(x)| \right).
\end{equation}
If this is not the case, then for all $t \in [x+ 1/2, x+ 3/2 ]$,  
\begin{equation} \label{Aeq:uatt}
|u(t) | <  \frac{C_3}{4}  \min \left(
\left\M \frac{1}{p} \right\M_{\tilde{I},{\rm loc}-}, 1 \right) \left( |u(x)| + |pu'(x)| \right).
\end{equation}
Using (\ref{Aeq:sum}), we conclude then that for all $t \in [x+ 1/2,
x+ 3/2 ]$,
\begin{eqnarray}
|pu'(t)|   & \geq & C_3 \left( |u(x)| + |pu'(x)| \right) - |u(t)| \\ 
 \mbox{ } &  > &  C_3 \left( |u(x)| + |pu'(x)| \right) \left( 1 - \frac{
    \min( \M \frac{1}{p} \M_{\tilde{I},{\rm loc}-}, 1)}{4} \right) \nonumber \\
  \mbox{ } & \geq & \frac{C_3}{2} \left( |u(x)| + |pu'(x)| \right), \nonumber
\end{eqnarray}
i.e., $pu'$ is strictly signed, and therefore
\begin{eqnarray} \label{Aeq:realval}
\frac{C_3}{2}  \min \left( \left\M \frac{1}{p} \right\M_{\tilde{I},{\rm loc}-}, 1
\right) \left( |u(x)| + |pu'(x)| \right) & > & |u(x+ 3/2 )|+|u(x +
1/2)| \nonumber \\ 
& \geq & \left| \int_{x +1/2 }^{x+3/2} \frac{1}{p(t)} pu'(t)dt \right|
 \\
& \geq & \frac{C_3}{2} \left( |u(x)| + |pu'(x)| \right) \int_{x+
  1/2}^{x+3/2} \frac{1}{p(t)}dt, \nonumber
\end{eqnarray}
which is an obvious contradiction. We have proven (\ref{Aeq:uatx0}).

Since the function $x \mapsto \int_a^x \frac{1}{p(t)}dt$ is continuous 
on $\tilde{I}$, it is uniformly continuous. Thus for $\epsilon >0$ defined by
the equation $8C_4 \epsilon:= C_3 \min \left( \left\M \frac{1}{p}
  \right\M_{\tilde{I},{\rm loc}-} ,1 \right)$ there
exists a $\delta > 0$ for which 
\begin{equation} \label{Aeq:nopeaks1}
\sup_{x \in I} \int_x^{x+ \delta} \frac{1}{p(t)}dt \, \leq \, \epsilon.
\end{equation}
In this case, we conclude that for any $|t-x_0| \leq \delta$,
\begin{eqnarray}
|u(t)-u(x_0)| & \leq & C_4  \left( |u(x)| + |pu'(x)| \right) \left| \int_{x_0}^t \frac{1}{p(s)}ds
\right| \\ \mbox{ } & \leq & \frac{C_3}{8}  \left( |u(x)| + |pu'(x)|
\right) \min \left( \left\M \frac{1}{p}
  \right\M_{\tilde{I},{\rm loc}-} ,1 \right). \nonumber
\end{eqnarray}
From this observation, (\ref{Aeq:intbd}) follows.
\end{proof}

For certain applications, we will need a variant of Lemma~\ref{Alem:L2b} which
is true for complex valued solutions of (\ref{Aeq:sleq}); note the argument in
(\ref{Aeq:realval}) fails if the solution is not real valued.

\begin{Lemma} \label{Alem:L2b2}
Let $p$ and $q$ be functions with $\frac{1}{p} \in L^{1}_{ {\rm loc}}(
\R)$, $p>0$ almost everywhere, and $q \in L^{1}_{ {\rm loc}}( \R)$. 
Suppose $u \neq 0$ is a solution of $-(pu')' + qu = 0$ which satisfies
$\mbox{Re}[u(x) \overline{pu'(x)}] = 0$ for some $x \in R$. Then,
there exists a constant $C$, depending only on the local $L^1$-norms
of $\frac{1}{p}$ and $q$, for which 
\begin{equation} \label{Aeq:intbd2}
\int_x^{x+2} |u(t)|^2 dt \geq C \left( |u(x)|^2 + | pu'(x)|^2 \right).
\end{equation}
\end{Lemma}

\begin{proof} It is enough to demonstrate (\ref{Aeq:intbd2}) for
  solutions which additionally satisfy
\begin{equation} \label{Aeq:gnorm}
|u(x)|^2 + |pu'(x)|^2 = 1,
\end{equation}
since for an arbitrary solution of (\ref{Aeq:sleq}) one may define the
normalized solution $\psi(t) := [|u(x)|^2 + |pu'(x)|^2 ]^{-1/2} \,
u(t)$ which does satisfy (\ref{Aeq:gnorm}). We continue as in the
proof of the previous lemma. 

For any solution of (\ref{Aeq:sleq}) which satisfies
(\ref{Aeq:gnorm}), there exists constants $0<C_1 <C_2 < \infty$ for
which
\begin{equation} \label{Aeq:globalbd}
C_1 \, \leq \, |u(t)|^2 \, + \, |pu'(t)|^2 \, \leq \, C_2, \ \ \ \
\mbox{for all } t \in [x,x+2],
\end{equation}
by Lemma~\ref{Alem:gb}. Let $0<a<1$ be given; we will choose such an
$a$ below. We claim that there exists an $x_0 \in [x,x+1]$ for which
\begin{equation} \label{Aeq:ubelow}
|u(x_0)| \, \geq \, a C_1.
\end{equation}
If we establish the existence of an $a$ and an $x_0$ for which
(\ref{Aeq:ubelow}) holds, then (\ref{Aeq:intbd2}) is true as
one may calculate that 
\begin{equation} \label{Aeq:moduder}
p(t) \frac{d}{dt} |u(t)|^2 \, = \, 2 \mbox{Re} \left[ \, u(t)
  \overline{pu'(t)} \, \right], 
\end{equation}
and therefore
\begin{eqnarray} \label{Aeq:diffinmod}
\left| \, |u(t)|^2 \, - \, |u(x_0)|^2 \, \right| & \leq & \int_{
  \min(t,x_0)}^{\max(t, x_0)} \frac{1}{p(s)} \, \left( \, |u(s)|^2 \, +
  \, |pu'(s)| \, \right) ds \\ \nonumber
& \leq & C_2 \, \int_{ \min(t,x_0)}^{\max(t, x_0)} \frac{1}{p(s)} \,
ds  \, \leq \, \frac{a C_1}{2}, \\ \nonumber
\end{eqnarray}
for $t$ sufficiently small. As in Lemma~\ref{Alem:L2b}, this
completes the proof.

To verify (\ref{Aeq:ubelow}), suppose it is not the case. Then
\begin{equation} \label{Aeq:usmall}
|u(t)|^2 \, < \, a C_1, \ \ \ \ \mbox{for all } t \in [x,x+1],
\end{equation} 
and hence (\ref{Aeq:globalbd}) implies that 
\begin{equation} \label{Aeq:dubig}
|pu'(t)|^2 \, > \, (1-a)C_1,  \ \ \ \ \mbox{for all } t \in [x,x+1],
\end{equation}
as well. Using (\ref{Aeq:moduder}), one may further calculate that
\begin{equation} \label{Aeq:moduderder}
\frac{d}{dt} \left( p(t) \frac{d}{dt} |u(t)|^2 \right) \, = \, 2
\left( \, \frac{|pu'(t)|^2}{p(t)} \, + \, \mbox{Re}[q(t)] \, |u(t)|^2 \, \right). 
\end{equation}

We can now estimate that
\begin{equation} \label{Aeq:start}
\int_x^{x+1} \frac{d}{dt} | u(t) |^2 \, dt  \, = \, |u(x+1)|^2 -
|u(x)|^2 \, < \, a C_1. 
\end{equation}
Moreover,
\begin{eqnarray}
\int_x^{x+1} \frac{d}{dt}  |u(t)|^2 \, dt & = & 
\int_x^{x+1} \frac{1}{p(t)} \int_x^t \frac{d}{ds} 
\left( \, p(s) \frac{d}{ds} |u(s)|^2 \,\right) ds \, dt \\ \nonumber
& =  & I_1 + I_2, 
\end{eqnarray}
where we have used (\ref{Aeq:moduder}), the boundary condition
$\mbox{Re}[u(x) \overline{pu'(x)}] = 0$, 
\begin{equation} \label{Aeq:int1}
I_1 \, = \, 2  \int_x^{x+1} \frac{1}{p(t)} \int_x^t \frac{ |pu'(s)|^2}{p(s)} \,
ds \, dt, 
\end{equation}
and
\begin{equation} \label{Aeq:int2}
I_2 \, = \, 2  \int_x^{x+1} \frac{1}{p(t)} \int_x^t \, \mbox{Re}[q(s)]
\, |u(s)|^2 \, ds \, dt. 
\end{equation}
With (\ref{Aeq:dubig}), it is clear that
\begin{equation}
I_1 \, \geq \, (1-a)C_1 \left( \int_x^{x+1} \frac{1}{p(t)} dt \, \right)^2.
\end{equation}
To bound $I_2$, we use (\ref{Aeq:usmall}) as follows
\begin{eqnarray}
I_2 & \geq &  - 2  \int_x^{x+1} \frac{1}{p(t)} \int_x^t \, \mbox{Re}[q(s)]_{-}
\, |u(s)|^2 \, ds \, dt \nonumber \\
 & \geq &  - 2 \, a \, C_1 \,  \int_x^{x+1} \frac{1}{p(t)} \int_x^t \, \mbox{Re}[q(s)]_{-}
\, ds \, dt, \label{Aeq:finish}
\end{eqnarray}
where for a real valued function $f$, $f_{-}(t) := \max(-f(t),
0)$. Collecting our bounds from (\ref{Aeq:start}) to (\ref{Aeq:finish}), we have proven
that
\begin{equation}
aC_1 \, > \, aC_1 \, \left[ \frac{1-a}{a} \left( \int_x^{x+1}
    \frac{1}{p(t)} dt \right)^2 \, - \, 2  \int_x^{x+1} \frac{1}{p(t)} \int_x^t \, \mbox{Re}[q(s)]_{-}
\, ds \, dt \, \right], 
\end{equation}
which is an obvious contradiction for $a$ sufficiently small. We have
established the claim (\ref{Aeq:ubelow}), and thus the lemma is proven.
\end{proof}

\begin{Remark} \label{rem:u=1} One added difficulty in the proof of
Lemma~\ref{Alem:L2b2} above, is the possibility that the 
solution $u$ may vanish at $x$. If one knows the complex valued
solution satisfies $|u(x)|^2 =1$, in contrast to $\mbox{Re}[u(x)
\overline{pu'(x)}] = 0$, then arguments as in (\ref{Aeq:diffinmod}) readily
provide lower bounds of the type found in (\ref{Aeq:intbd2}).
\end{Remark}

\setcounter{equation}{0}


\section{The m-function} \label{App:mfun}

The goal of this section is to prove Theorem~\ref{Bthm:mc} below which
concerns the compact uniform convergence of $m$-functions corresponding to a sequence
of Sturm-Liouville equations whose coefficients converge in 
$L^1_{{\rm loc}}(\R)$. For a more indepth discussion of $m$-function
theory, we refer the reader to Chapter 9 of \cite{CL}, and for
convenience, we adopt the notation used therein. 

Let $\{ p_n \}_{n=0}^{\infty}$ and $\{ q_n \}_{n=0}^{\infty}$ be sequences of real-valued functions
which satisfy $p_n >0$, $\frac{1}{p_n} \in L^1_{{\rm loc}}( \R)$, and 
$q_n \in L^1_{{\rm loc}}( \R)$ for all $n \geq 0$. Let $K \subset
\C^+$, the complex upper half plane, be compact and
consider solutions of 
\begin{equation} \label{Beq:sleqn}
-(p_n u')' + q_n u = \lambda u
\end{equation}
for $\lambda \in K$. For $n \geq 0$ and $x \in \R$, let $\phi_{n,x}$ and
$\theta_{n,x}$ be the solutions of (\ref{Beq:sleqn})
which satisfy the boundary conditions
\begin{equation} \label{Beq:bcs}
\left( \begin{array}{cc} \phi_{n,x}(x, \lambda) & \theta_{n,x}(x, \lambda) \\
p \phi_{n,x}'(x, \lambda) & p \theta_{n,x}'(x, \lambda) \end{array} \right) \, = \, I.
\end{equation} 
Given any $y>x$ denote by  
\begin{equation} \label{Beq:disc}
D_{n, x, \lambda}^y \, := \, \left\{ m \in \C^+ : \int_x^y | \phi_{n,x}(t,
\lambda) + m \theta_{n,x}(t, \lambda)|^2dt \, \leq \, \frac{
\mbox{Im}[m]}{\mbox{Im}[ \lambda]} \right\}
\end{equation}
the disc of radius $r_{n,x, \lambda}^y$ and center $c_{n,x,
  \lambda}^y$ in the upper half plane. In the limit as $y \to \infty$,
the boundary of these discs is given by the solutions, $m_n(x,
\lambda)$, of the equation
\begin{equation} \label{Beq:meq}
\int_x^{\infty}| \phi_{n,x}(t, \lambda) +m_n(x, \lambda)\theta_{n,x}(t,
\lambda)|^2dt \, = \, \frac{ \mbox{Im}[m_n(x,
\lambda)]}{\mbox{Im}[\lambda]}.
\end{equation}

\begin{Lemma} \label{Blem:mbds} Let $\{ p_n \}_{n=0}^{\infty}$ 
and $\{ q_n \}_{n=0}^{\infty}$ be sequences of real-valued functions
which satisfy $p_n >0$, $\frac{1}{p_n} \in L^1_{{\rm loc}}( \R)$, and 
$q_n \in L^1_{{\rm loc}}( \R)$ for all $n \geq 0$. Fix 
$I=[a,b] \subset \R$ and $K \subset \C^+$ compact. If 
$ \frac{1}{p_n} \to \frac{1}{p_0}$ and $ q_n \to q_0$ in $L^1_{ {\rm
    loc}}( \R)$, then the union of discs  
\begin{equation} \label{Beq:uDs}
D : = \bigcup_{ \lambda \in K, n \geq 0, x \in I} D_{n, x,
\lambda}^{b+2} 
\end{equation}
is a bounded subset of $\C^+$. Moreover, every solution 
$m_n(x, \lambda)$ of (\ref{Beq:meq}) satisfies 
\begin{equation} \label{Beq:mlb}
{\rm Im}[m_n(x, \lambda)] \, \geq \, C \, > \,
0, 
\end{equation}
where $C$ may be choosen uniform in the parameters: $n \geq 0$, $x \in
I$, and $\lambda \in K$. 
\end{Lemma}
\begin{proof} 
We will prove that the union of discs $D$ is bounded by deriving
uniform estimates on the center and radius of each 
$D_{n, x, \lambda}^{b+2}$. Recall from \cite{CL} that
\begin{equation} \label{Beq:Dc2}
c_{n, x, \lambda}^{b+2} \, = \, - \frac{ [\phi_{n,x}, \theta_{n,x}]( b+2 )} {2i \mbox{Im}[
\lambda] \cdot \int_x^{b+2} |\theta_{n,x}(t, \lambda)|^2dt},
\end{equation}
and
\begin{equation} \label{Beq:Dr2}
r_{n, x, \lambda}^{b+2} \, = \, \frac{1}{2 \mbox{Im}[ \lambda] \cdot \int_x
^{b+2} |\theta_{n,x}(t, \lambda)|^2dt},
\end{equation}
where $ [\phi_{n,x}, \theta_{n,x}](y) = \phi_{n,x}(y) \overline{p_n \theta_{n,x}'}(y) -
  p_n \phi_{n,x}'(y)\overline{\theta_{n,x}}(y) $.  
With $\tilde{I} = [a,b+2]$, the $L^1_{ {\rm loc}}$-convergence of the
coefficients of (\ref{Beq:sleqn}) implies that:
\begin{equation} \label{Beq:uni+s}
\max \left\{ \sup_{n \geq 0} \left\| \frac{1}{p_n} \right\|_{ \tilde{I},+}, \quad
  \sup_{n \geq 0} \sup_{\lambda \in K} \left\| q_n - \lambda
  \right\|_{ \tilde{I},+} \right\} \, < \, \infty,
\end{equation}
\begin{equation} \label{Beq:uni-s}
\inf_{n \geq 0} \left\| \frac{1}{p_n} \right\|_{ \tilde{I},-} \, > \, 0,
\end{equation}
and for every $\epsilon >0$, there exists $\delta >0$ for which
\begin{equation} \label{Beq:uninopeaks}
\sup_{n \geq 0} \sup_{x \in I} \int_x^{x+ \delta} \frac{1}{p_n(t)} dt
\, \leq \, \epsilon.
\end{equation}
Inserting (\ref{Beq:uni+s}), (\ref{Beq:uni-s}), and (\ref{Beq:uninopeaks}) into the proof of
Lemma~\ref{Alem:L2b2}, one estimates that
\begin{eqnarray}
\int_x^{b+2} | \theta_{n,x}(t, \lambda) |^2 dt & \geq & \int_x^{x+2} |
\theta_{n,x}(t, \lambda) |^2 dt \nonumber \\
\mbox{ } & \geq & C \left(  | \theta_{n,x}(x, \lambda) |^2 +  | p_n
  \theta_{n,x}'(x, \lambda) |^2 \right) \nonumber \\
\mbox{ } & = & C,
\end{eqnarray}
with a constant $C>0$ which is uniform in the parameters: $x \in I$, 
$n \geq 0$, and $\lambda \in K$. From this bound, it is clear that
\begin{equation} \label{Beq:rbd}
| r_{n, x, \lambda}^{b+2}| \, \leq \, \inf_{ \lambda \in
  K} \frac{ (2C)^{-1}}{ {\rm Im}[ \lambda]},
\end{equation}
and
\begin{equation} \label{Beq:cbd}
| c_{n, x, \lambda}^{b+2}| \, \leq \, \inf_{ \lambda \in
  K} \frac{ C'}{ {\rm Im}[ \lambda]},
\end{equation}
where we note that the constant $C'$, appearing in (\ref{Beq:cbd}), 
incorporates a repeated application of Lemma~\ref{Alem:gb}
to bound the generalized Wronskian $[\phi_{n,x}, \theta_{n,x}](b+2)$.
This proves the boundedness of $D$.

Similarly, using (\ref{Beq:disc}) and Remark~\ref{rem:u=1}, one sees that any point 
$m \in D_{n, x, \lambda}^{b+2}$ satisfies 
\begin{eqnarray}
{\rm Im} [ m  ]  & \geq & {\rm Im}[ \lambda ]
\int_x^{x+2} | \phi_{n,x}(t, \lambda) + m \theta_{n,x}(t, \lambda) |^2 dt  \nonumber \\
\mbox{ } & \geq &  {\rm Im}[ \lambda] \, C \, \left( 1+ |m|^2
\right) \nonumber \\
\mbox{ } & \geq & C \, \inf_{ \lambda \in K} {\rm Im}[ \lambda],
\end{eqnarray}
which proves (\ref{Beq:mlb}).
\end{proof}

\begin{Theorem} \label{Bthm:mc} Let $\{ p_n \}_{n=0}^{\infty}$ 
and $\{ q_n \}_{n=0}^{\infty}$ be sequences of real-valued functions
which satisfy $p_n >0$, $\frac{1}{p_n} \in L^1_{{\rm loc}}( \R)$, and 
$q_n \in L^1_{{\rm loc}}( \R)$ for all $n \geq 0$. Fix 
$I=[a,b] \subset \R$ and $K \subset \C^+$ compact. If 
$ \frac{1}{p_n} \to \frac{1}{p_0}$ and $ q_n \to q_0$ in $L^1_{ {\rm
    loc}}( \R)$ and the equation 
\begin{equation} \label{Beq:lpeq}
 -(p_0u')' + q_0u= \lambda u 
\end{equation}
is limit point at $ + \infty $, then 
\begin{equation} \label{Beq:mconv}
m_n(x, \lambda) \rightarrow m_0(x, \lambda), 
\end{equation}
uniformly for $(x, \lambda) \in I \times K$.  
\end{Theorem}

\begin{proof}
To prove this theorem, we will first establish pointwise convergence.
Fix $x_0 \in I$ and $\lambda_0 \in K$. As (\ref{Beq:lpeq}) is limit point at
$+ \infty$, the disc $D_{0, x_0, \lambda_0}^y$ shrinks to a unique
point in the limit $ y \to \infty$. Thus, by taking $y$
large, we can make the radius $r_{0, x_0, \lambda_0}^y$
arbitrarily small. For $y'$ large but fixed, Lemma~\ref{Alem:db} implies
that $r_{n, x_0, \lambda_0}^{ y'} \to r_{0, x_0, \lambda_0}^{y'}$ as
the functions $\theta_{n, x_0}$ all satisfy the same
normalization. From this we conclude that for all sufficiently large
$n$, the radii $r_{n, x_0, \lambda_0}^{ y'}$ can be made
arbitrarily small as well. Moreover, since 
$m_0(x_0, \lambda_0)$ is in the interior of 
$D_{0, x_0, \lambda_0}^{ y'}$ for all $y' > x_0$, we see from (\ref{Beq:disc}) that 
\begin{equation} \label{Beq:m0in}
\int_{x_0}^{ y'}| \phi_{0, x_0}(t, \lambda_0)+m_0(x_0,\lambda_0)
\theta_{0, x_0}(t,
\lambda_0)|^2dt < \frac{ \mbox{Im}[m_0(x_0,
\lambda_0)]}{\mbox{Im}[\lambda_0]}.
\end{equation} 
Appealing again to Lemma \ref{Alem:db}, we conclude that 
for $n$ sufficiently large,
\begin{equation} \label{Beq:mjin}
\int_{x_0}^{ y'}| \phi_{n, x_0}(t,
\lambda_0)+m_0( x_0 ,\lambda_0) \theta_{n, x_0}(t, \lambda_0)|^2dt <
\frac{ \mbox{Im}[m_0( x_0, \lambda_0)]}{\mbox{Im}[\lambda_0]},
\end{equation}
as well. In other words, for large enough $n$, $m_0(x_0,\lambda_0) \in
D_{n, x_0, \lambda_0}^{ y'}$, and therefore, $|m_n( x_0, \lambda_0)-m_0(x_0,
\lambda_0)| \leq 2 \cdot r_{n, x_0, \lambda_0}^{ y'}$. As we have
already argued, these radii become arbitrarily small, and therefore we have 
proven pointwise convergence. In fact, as the functions 
$m_n( x_0, \cdot)$ are analytic in $\C^+$, the uniform bounds 
proven in Lemma~\ref{Blem:mbds} combined with Montel's Theorem, see
e.g. \cite{Stein}, imply that the pointwise convergence is uniform
for $\lambda \in K$. 

The full result now follows from the Ricatti equation, i.e., the fact that 
\begin{equation} \label{Beq:ric}
\partial_x m_n(x,\lambda) = q_n(x) - \lambda - \frac{1}{p_n(x)}m_n(x, \lambda)^2.
\end{equation}
For any $a \leq t \leq b$, integration of (\ref{Beq:ric})
yields the following estimate:
\begin{multline} \label{Beq:mdiff}
|m_n(t, \lambda) - m_0(t, \lambda)| \, \leq \, | m_n(a, \lambda)-m_0(a, \lambda)| +
\int_a^t|q_n(s)-q_0(s)|ds + \\
+ \int_a^t \left| \frac{1}{p_n(s)}-\frac{1}{p_0(s)} \right| \cdot |m_0(s,
\lambda)|^2 ds + \int_a^t \frac{1}{p_n(s)} \cdot |m_n(s, \lambda)^2
- m_0(s, \lambda)^2|ds.
\end{multline}
As each $m_n(s, \lambda) \in D_{n, s, \lambda}^{b+2}$, Lemma
\ref{Blem:mbds} implies that there exists $M>0$ for which 
$|m_n(s, \lambda)| \leq M$ uniformly in the parameters 
$n \geq 0$, $s \in I$, and $\lambda \in K$. Inserting this upper bound
into (\ref{Beq:mdiff}) implies that  
\begin{multline}
|m_n(t, \lambda) - m_0(t, \lambda)| \, \leq \, | m_n(a, \lambda)-m_0(a,
\lambda)| + \M q_n -q_0 \M_I  \\ 
+ M^2 \cdot \left\M \frac{1}{p_n} - \frac{1}{p_0} \right\M_I  + 2M \cdot
\int_a^t \frac{1}{p_n(s)} \cdot |m_n(s, \lambda) - m_0(s, \lambda)|ds.
\end{multline}
Applying Gronwall's Lemma, see e.g. \cite{Walter}, we find that for any $t \in I$,
\begin{equation} \label{Ceq:gron}
|m_n(t, \lambda) - m_0(t, \lambda)| \, \leq \, C(I, \lambda,n) \cdot \exp \left( 2M \cdot
  \sup_{n \geq 0} \left\M   \frac{1}{p_n} \right\M_I \right), 
\end{equation}
where 
\begin{equation} \label{Ceq:feq}
C(I,\lambda, n) \, := \, | m_n(a, \lambda)-m_0(a, \lambda)| \, + \, \M q_n -q_0 \M_I
\, + \, M^2 \cdot \left\M \frac{1}{p_n} - \frac{1}{p_0} \right\M_I.
\end{equation}
The pointwise result we proved above demonstrates that $C(I, \lambda, n)
\rightarrow 0$ uniformly for $\lambda \in K$. We have proven the
theorem. 
\end{proof}

\setcounter{equation}{0}


\section{m-Asymptotics and Solution Estimates} \label{App:mfunasy}

We may now reproduce certain estimates found in \cite{Ben1}.
We will consider the equation
\begin{equation} \label{Ceq:bst}
-(pu')' +qu = \lambda u,
\end{equation}
where $p$ and $q$ are real valued functions with $p>0$ almost
everywhere, both $\frac{1}{p}$ and $q$ are locally integrable, and
$\lambda \in \C$. We impose a further condition on the 
coefficients, namely that (\ref{Ceq:bst}) is
limit point at $+ \infty$. The crux of 
Bennewitz's argument is a clever rescaling of the
coefficients in (\ref{Ceq:bst}) as the energy parameter $\lambda$
varies along a ray in the complex plane. From this, he derives an asymptotic formula for
the $m$-function and a related result for the Weyl solution.


\subsection{The Scaling}
Fix $x \in \R$ and denote by $P_x : [x,
\infty) \to [0, \int_x^{\infty} \frac{1}{p(s)}ds)$ and $W_x : [x,
\infty) \to [0, \infty)$ the functions defined by 
\begin{equation} \label{Ceq:scale}
\begin{array}{ccc} P_x(t):= \int_x^t
\frac{1}{p(s)}ds & \mbox{and} & W_x(t):=t-x. 
\end{array}
\end{equation}
As both $P_x$ and $W_x$ are continuous and strictly increasing, we may
define their inverses $P_x^{-1}$ and $W_x^{-1}$, respectively, each of
which is also continuous and strictly increasing. Consider the
function $\tilde{f}_x: (0, \int_x^{ \infty} \frac{1}{p(s)} ds)
\rightarrow (0, \infty)$ defined by 
\begin{equation} \label{Ceq:tf}
 \tilde{f}_x(t) \, = \,\frac{1}{t \, W_x \left( P^{-1}_x(t) \right)
 } \, = \, \frac{1}{t \left( P_x^{-1}(t) -x \right)}. 
\end{equation} 
Observe that $\tilde{f}_x$ is continuous and strictly decreasing with $\lim_{t \rightarrow 0^+}
\tilde{f}_x(t) = \infty$ and  $\tilde{f}_x(t)  \to 0$ as $t \to
\int_x^{ \infty} \frac{1}{p(s)}ds$. In this case, we set $f_x:=\tilde{f}_x^{-1}$.
\begin{Lemma} \label{Clem:fscale} For fixed $x \in \R$, we have that
\begin{equation} \label{Ceq:fasy}
\lim_{r \rightarrow \infty}f_x(r) = 0, \ \ \mbox{while} \ \
\lim_{r \rightarrow \infty}r f_x(r) = \infty.
\end{equation} Moreover, for any pair of numbers $(r,t)$ with
$r>0$ and $t>x$, 
\begin{equation} 
 r \, P_x(t) \, W_x(t) =1 \quad \mbox{if and only if} \quad r \, f_x(r) \, W_x(t) =1.
\end{equation} 
In particular, for such a pair $(r,t)$,  $f_x(r)=P_x(t)$.
\end{Lemma}
\begin{proof} Let $x \in \R$ be fixed. As $\tilde{f}_x$ is invertible, 
for each $r>0$ there exists a unique $t_r>0$ for which 
$r = \tilde{f}_x(t_r)$. Using the above observations, as
$r \rightarrow \infty$,  $t_r \rightarrow 0$ proving the first claim
in (\ref{Ceq:fasy}). Direct substitution shows that 
\begin{equation} 
rf_x(r) = \frac{1}{W_x \left( P^{-1}_x(t_r) \right)},
\end{equation}
from which the later portion of (\ref{Ceq:fasy}) is clear. 

Next, suppose that $r>0$ and $t>x$ are choosen to satisfy the equation 
$r \, P_x(t) \, W_x(t)=1$. In this case, 
\begin{equation} \label{Ceq:unrav}
f_x(r) = f_x \left( \frac{1}{P_x(t) \cdot W_x
\left(P^{-1}_x(P_x(t))\right)} \right)= P_x(t) =
\frac{1}{rW_x(t)}.
\end{equation}
Similarly, if $r \, f_x(r) \, W_x(t) =
1$, then $f_x(r) = \frac{1}{rW_x(t)}$ and therefore $r = \tilde{f}_x(
\frac{1}{rW_x(t)})$. Rewriting things, one sees that 
$W_x(t)=W_x(P^{-1}_x(\frac{1}{rW_x(t)}))$, which implies that $t = P_x^{-1}(
\frac{1}{rW_x(t)})$, and hence $r \, P_x(t) \, W_x(t)=1$.
We have proven the lemma. 
\end{proof}


\subsection{m-function asymptotics}
Fix $x \in \R$ and let $\phi_x$ and
$\theta_x$ be the solutions of (\ref{Ceq:bst})
which satisfy the boundary conditions
\begin{equation} \label{Ceq:bcs}
\left( \begin{array}{cc} \phi_x(x, \lambda) & \theta_x(x, \lambda) \\
p \phi_x'(x, \lambda) & p \theta_x'(x, \lambda) \end{array} \right) \, = \, I.
\end{equation} 
The following theorem is proven in \cite{CL}:
\begin{Theorem} \label{Cthm:Weyl}
Let $\phi_x$ and $\theta_x$ be the solutions of (\ref{Ceq:bst}),
corresponding to $\lambda \in \C^+$, which
satisfy the boundary conditions given by (\ref{Ceq:bcs}).
The linear combination $\psi_x = \phi_x +m \theta_x$ has 
the property that $m$ is the $m$-function, $m(x, \lambda)$, of (\ref{Ceq:bst})  
if and only if $\lim_{t \rightarrow \infty} [ \psi_x,\psi_x](t) =0$, where
\begin{equation} \label{Ceq:Wron}
[f,g](t) \, := \, f(t) \overline{pg'}(t) -
  pf'(t)\overline{g}(t) 
\end{equation}
is the modified Wronskian corresponding to (\ref{Ceq:bst}). 
\end{Theorem}
\noindent This theorem is a direct consequence of the equation
\begin{equation}
[ \psi_x, \psi_x](y) = 2 i {\rm Im}[ \lambda] \int_x^y |
\psi_x|^2 dt - 2 i {\rm Im}[m]
\end{equation}
which is used to define the discs $D_{x, \lambda}^y$
introduced previously in (\ref{Beq:disc}); we refer to \cite{CL, EasthamKalf} for the details. 

We will now consider the properties of solutions of equation (\ref{Ceq:bst}) 
as the energy parameter is varied along a ray in the complex upper
half plane. Fix $\mu \in \C^+$, take $r>0$, and suppose $u(t, r \mu)$ is a
solution of (\ref{Ceq:bst}) corresponding to $\lambda = r \mu$. For fixed $x \in
\R$ set $s:= \frac{1}{rf_x(r)} $ where $f_x$ is as defined above.
Denote by 
\begin{equation} \label{Ceq:newcoef}
q_r(t) \, := \, \frac{q(st+x)}{r}, \quad \frac{1}{p_r(t)} \,
:= \,  \frac{s}{f_x(r)p(st+x)},
\end{equation} and $u_r(t):=u(st+x, r \mu)$. A short
calculation shows that 
\begin{equation} \label{Ceq:trans}
- \frac{d}{dt} \left( p_r \frac{d}{dt}u_r
\right) + (q_r- \mu)u_r \, = \, \frac{1}{r} \left[ -(pu')'+(q-r \mu)u
\right], 
\end{equation}
and we see that if $u$ solves (\ref{Ceq:bst}), then $u_r$ solves
\begin{equation} \label{Ceq:mod}
- \frac{d}{dt} \left( p_r \frac{d}{dt}u_r
\right) + q_ru_r =  \mu u_r.
\end{equation}
Take $\theta_r(t):= \frac{\theta_x(st+x, r \mu)}{f_x(r)}$ and 
$\phi_r(t):= \phi_x(st+x, r \mu)$. Observe that both $\theta_r$ and 
$\phi_r$ are solutions of the modified equation (\ref{Ceq:mod}) which
satisfy the boundary conditions
\begin{equation} \label{Ceq:newbc}
\left( \begin{array}{cc} \phi_r(0) & \theta_r(0) \\ p_r \dot{\phi}_r(0) &
p_r \dot{\theta}_r(0) \end{array} \right) \, = \, I,
\end{equation}
where we have used the notation $\dot{f} := \frac{d}{dt}f$. 
\begin{Lemma} If $m(x, r \mu)$ is the $m$-function corresponding to
(\ref{Ceq:bst}) on $[x, \infty)$, then
\begin{equation} \label{Ceq:modm}
m_r(0, \mu) \, := \, f_x(r)m(x,r \mu), 
\end{equation}
is the $m$-function for (\ref{Ceq:mod}) on $[0, \infty)$.
\end{Lemma}
\begin{proof}
Let $\psi_r(t):=\phi_r(t)+m_r(0, \mu) \theta_r(t)$, where $m_r(0,
\mu)$ is as defined in (\ref{Ceq:modm}). One easily verifies that
\begin{equation} \label{Ceqn:simWron}
[\psi_r,\psi_r](t) = f_x(r) [\psi_x, \psi_x](ts+x),
\end{equation} 
where, as before, $\psi_x = \phi_x + m(x, r \mu) \theta_x$. 
Using Theorem \ref{Cthm:Weyl}, if $m(x, r \mu)$ is the 
$m$-function corresponding to (\ref{Ceq:bst}) with $\lambda = r \mu$, 
then (\ref{Ceqn:simWron}) implies that $\lim_{t \to \infty} [\psi_r,\psi_r](t)
= 0$, and therefore the lemma follows from another application of Theorem \ref{Cthm:Weyl}. 
\end{proof}
\begin{Theorem} \label{Cthm:masyt}
Suppose $p$ and $q$ are real valued functions for which $p>0$,
$\frac{1}{p} \in L^1_{loc}(\R)$, $q \in L^1_{loc}(\R)$, and equation
(\ref{Ceq:bst}) is limit point at $\infty$. Assume
in addition that x is a Lebesgue point of $\frac{1}{p}$. Then, as
$r \rightarrow \infty$,
\begin{equation} \label{Ceq:masy}
m(x, r \mu) = i \sqrt{r \mu} \cdot \sqrt{p(x)} + o( \sqrt{r}),
\end{equation}
where the square root above is the principle branch and the
convergence is uniform for $\mu$ in compact subsets of $\C^+$. 
\end{Theorem}
\begin{proof} Fix an arbitrary $c>0$. One may calculate that
\begin{equation} \label{Ceq:qint} 
\int_0^c |q_r(t)| \, dt \, = \, \frac{1}{r} \int_0^c |q(st+x)| \, dt \, 
= \, f_x(r) \int_x^{cs+x} |q(y)| \, dy.
\end{equation} 
By Lemma \ref{Clem:fscale}, both $s$ and $f_x(r)$ go to zero as $r$ goes to
infinity, and so the same is true for the above integral.
A similar result holds for $\frac{1}{p_r}$. To see this, set 
$\tilde{s} := W_x^{-1}(s)$. Clearly then, 
\begin{equation} \label{Ceq:sscale}
 r \, f_x(r) \, W_x( \tilde{s}) \, = \, r \, f_x(r) \, s \, = \, 1,
\end{equation}
and from Lemma \ref{Clem:fscale} we may conclude that $f_x(r) = P_x(
\tilde{s})$. In this case, one may calculate that
\begin{equation} \label{Ceq:pint}
 \int_0^c \frac{1}{p_r(t)}dt  \,  = \, 
\frac{s}{\, f_x(r)} \int_0^c \frac{1}{p(ts+x)}dt  \, = \,  
\frac{W_x( \tilde{s})}{P_x( \tilde{s})} \, \frac{P_x(cs+x)}{cs} \, c.
\end{equation}
Observe that as $r \rightarrow \infty$, $\tilde{s} \rightarrow x$. 
Thus, since $x$ is a Lebegue point of $\frac{1}{p}$, the product of the ratios 
$ \frac{W_x( \tilde{s})}{P_x( \tilde{s})}$ and $\frac{P_x(cs+x)}{cs}$
goes to one as $r \rightarrow \infty$. These calculations show 
that as $r \rightarrow \infty$, $q_r \rightarrow 0$ and $\frac{1}{p_r} \rightarrow 1$ in $L^1_{loc}(0,
\infty)$. Using Theorem \ref{Bthm:mc}, we may conclude that 
\begin{equation} \label{Ceq:fm}\lim_{r \rightarrow \infty}m_r(0, \mu) = i \sqrt{ \mu}, 
\end{equation}
which is the well-known $m$-function for the free equation. 

Applying Lemma \ref{Clem:fscale} again, (\ref{Ceq:sscale}) 
implies not only that $f_x(r) = P_x( \tilde{s})$, but also 
$r \, P_x( \tilde{s}) \, W_x( \tilde{s})=1$.
From these equations it is easy to see that 
\begin{equation} \label{Ceq:sqrf} 
\sqrt{r}f_x(r) \, = \, \sqrt{ \frac{1}{P_x( \tilde{s}) \,
W_x( \tilde{s})}} \, P_x( \tilde{s}) \, = \, \sqrt{ \frac{P_x(
\tilde{s})}{W_x( \tilde{s})}},
\end{equation}
and thus 
\begin{equation}
 \lim_{r \rightarrow \infty} \sqrt{r}f_x(r) \, = \,
\lim_{\tilde{s} \rightarrow x} \sqrt{ \frac{P_x(
\tilde{s})}{W_x(\tilde{s})}} \, = \, \sqrt{ \frac{1}{p(x)}}, 
\end{equation}
where again we have used that $x$ is a Lebesgue point of
$\frac{1}{p}$. Clearly then,
\begin{equation} \label{Ceq:masy2}
 \lim_{r \rightarrow \infty} \frac{m(x,r \mu)}{\sqrt{r}} \, = \,
 \lim_{r \rightarrow \infty} \frac{ m_r(0, \mu)}{ \sqrt{r}f_x(r)} = i
\sqrt{\mu} \cdot \sqrt{p(x)},
\end{equation}
and we have proven (\ref{Ceq:masy}).
\end{proof}


\subsection{The growth of solutions}
For any $x \in \R$ and $\lambda \in \C^+$, the $m$-function may be written as
\begin{equation} \label{Ceq:m=psi}
 m(x, \lambda) = \frac{p \psi'(x,
\lambda)}{\psi(x, \lambda)},
\end{equation} 
where $\psi$ is the Weyl solution corresponding to (\ref{Ceq:bst}).
Upon integration one finds that
\begin{equation} \label{Ceq:psi1}
\psi(x, \lambda) \, = \,  \psi(0,
\lambda) \, + \, \int_0^x \frac{1}{p(t)}m(t, \lambda) \psi(t,
\lambda)dt, 
\end{equation}
and therefore,
\begin{equation} \label{Ceq:epsi}
 \psi(x, \lambda) \, = \, \psi(0, \lambda) \, \exp \left( \int_0^x \frac{1}{p(t)}m(t, \lambda)
   dt \right).
\end{equation}
We may now state 

\begin{Theorem} \label{Cthm:solasy} Let $I:=[0,b]$ and $K \subset
  \C^+$ compact be fixed. For any $\mu \in K$ and $r>0$, let $\psi$ be
  the Weyl solution of (\ref{Ceq:bst}) corresponding to $\lambda = r
  \mu$. One has that
\begin{equation} \label{Ceq:solasy}
\lim_{r \rightarrow \infty} \frac{1}{ i \sqrt{r}} \, \ln \left[
  \frac{\psi(x,r \mu)}{ \psi(0, r \mu)} \right] \, = \, \int_0^x \sqrt{ \frac{ \mu}{ p(t)}}dt,
\end{equation}
where the convergence is uniform for $(x, \lambda) \in I \times K$. 
\end{Theorem}

Formally, equation (\ref{Ceq:solasy}) follows readily by combining 
(\ref{Ceq:masy}) and (\ref{Ceq:epsi}) above. Justifying the use 
of dominated convergence, however, requires a bit of work. To 
prove this theorem, we use a lemma, due to Hardy, concerning maximal functions.

Let $\mu_1$ and $\mu_2$ are two non-negative measures on $\R$
which are both absolutely continuous with respect to Lebesgue measure.
Fix an open interval $I \subset \R$, and define for any $t \in I$
\begin{equation}
\mu(t) : = \sup_{ \stackrel{x,y \in I:}{x<t<y}} \frac{ \mu_1[(x,y)]}{ \mu_2[(x,y)]}.
\end{equation}

\begin{Lemma} \label{Clem:meas} Suppose $\mu_1[I] < \infty$, then
\begin{equation}
\mu_2[ \{t \in I : \mu(t) > s >0 \}] \leq \frac{4 \mu_1[I]}{s}.
\end{equation}
\end{Lemma}

A nice proof of this lemma appears in the appendix of \cite{Ben1}.    

\noindent {\em Proof of Theorem \ref{Cthm:solasy}}
\newline Rewriting (\ref{Ceq:epsi}) yields  
\begin{equation} \label{Ceq:sol=m}
\frac{1}{i \sqrt{r}} \, \ln \left[ \frac{ \psi(x,r \mu)}{ \psi(0, r \mu)} \right] \, = \, \int_0^x
\frac{1}{p(t)} \frac{m(t, r \mu)}{i \sqrt{r}} \, dt \, = \, \int_0^x
\frac{1}{p(t)} \frac{m_r(0, \mu)}{i \sqrt{r}f_t(r)} \, dt,
\end{equation}
where, for the last equality above, we have used (\ref{Ceq:modm}) and the
quantities defined in the previous subsections. 
Define the following non-negative measure on $\R$ 
\begin{equation} \label{Ceq:mu2}
\mu_2[(a,b)] \, := \, \int_a^b \frac{1}{p(t)}dt,
\end{equation} 
and note that (\ref{Ceq:sol=m}) can be rewritten as
\begin{equation} \label{Ceq:sol=mu2}
\frac{1}{i \sqrt{r}} \,  \ln \left[ \frac{ \psi(x,r \mu)}{ \psi(0, r \mu)} \right] \, = \, \int_0^x \frac{m_r(0,
\mu)}{i \sqrt{r}f_t(r)} \, d\mu_2(t).
\end{equation}
Using (\ref{Ceq:sqrf}) and the fact that the convergence in
(\ref{Ceq:fm}) is uniform with respect to $\mu \in K$, 
as proven in Theorem~\ref{Bthm:mc}, we see that there exists $C>0$ for
which 
\begin{equation} \label{Ceq:bound} 
\left| \frac{m_r(0, \mu)}{i \sqrt{r}f_t(r)} \right| \, \leq \, 
C \, \sqrt{ \frac{W_t(\tilde{s})}{P_t(\tilde{s})}},
\end{equation}
if $r$ sufficiently large. Here $C=C(K)$. Now set
\begin{equation} \label{Ceq:g}
g(t) \, := \, \sup_{ y \in (t, 2b)} \frac{W_t(y)}{P_t(y)}.
\end{equation}

We now claim that for $r$ sufficiently large,
\begin{equation}  \label{Ceq:bd2} 
\left| \frac{m_r(0, \mu)}{i \sqrt{r}f_t(r)} \right| \, \leq \, 
C \, \sqrt{g(t)}.
\end{equation}
Since $I$ is compact and the function $P(y) := \int_0^y \frac{1}{p(t)}
\, dt$ is continuous, the number $P_- := \inf_{y \in I}
P(y+b) -P(y)$ is strictly positive. If $r$ is chosen such that $r
P_- b \geq 1$, then, as in the proof of Theorem~\ref{Cthm:masyt}, with $s =
\frac{1}{r f_t(r)}$ and $\tilde{s} = W_t^{-1}(s)$ we have that $r P_t(
\tilde{s}) W_t( \tilde{s}) = 1$. Thus,  
\begin{equation}
\int_t^{\tilde{s}} \frac{1}{p(y)} \, dy \, ( \tilde{s} - t) \, = \,
P_t( \tilde{s}) W_t( \tilde{s}) \, = \frac{1}{r} \, \leq \, P_- b,
\end{equation}
from which it is clear that $\tilde{s} \leq b + t \leq 2b$. This
proves (\ref{Ceq:bd2}). 

We are now ready to apply Lemma~\ref{Clem:meas}. Let $\tilde{I} = (0,
2b)$ be the open interval under consideration, let $\mu_1$ be defined
by $\mu_1[(a,b)] = b-a$, and let $\mu_2$ be as defined in
(\ref{Ceq:mu2}). Clearly, $\mu(t) \geq g(t)$, i.e. the two-sided
maximal function $\mu$ dominates the one-sided maximal function $g$. 
From this inequality, the following set containment is obvious: 
$A_s := \{ t \in I : g(t) >s >0 \} \subset \tilde{A}_s := \{ t \in
\tilde{I} : g(t) > s >0 \} \subset \{ t \in \tilde{I} : \mu(t) > s >0
\}$, and therefore,  
\begin{equation} \label{Ceq:meabound}
\mu_2[ A_s ] \, \leq \, \mu_2[ \tilde{A}_s ] \, \leq \, \mu_2[ \{t \in \tilde{I}: \mu(t)>s>0 \} ] \leq \frac{8b}{s}.
\end{equation} 
Hence, for any $x \in I$ one may estimate 
\begin{eqnarray} \label{Ceq:gest}
 \int_0^x \sqrt{g(t)}d \mu_2(t) &  \leq & \frac{1}{2} \int_0^1 s^{-1/2} \mu_2[ A_s]ds + \frac{1}{2} \int_1^{\infty} s^{-1/2}
\mu_2[ A_s]ds \nonumber \\ \mbox{ } & \leq & \mu_2[I] + 8b.  
\end{eqnarray}
Using the bounds in (\ref{Ceq:bd2}) and (\ref{Ceq:gest}), we are
justified in applying dominated convergence to (\ref{Ceq:sol=mu2}), and the
theorem is proven. \hfill \qed

\end{appendix}

\end{document}